\documentclass[onecolumn]{aastex6}

\usepackage{amsmath}
\usepackage{hyperref}
\usepackage{gensymb}
\usepackage{color}
\usepackage{soul}
\usepackage[normalem]{ulem}
\usepackage[mathscr]{euscript}

\def\be{\begin{equation}}
\def\ee{\end{equation}}
\def\ba{\begin{array}}
\def\ea{\end{array}}
\def\bea{\begin{eqnarray}}
\def\eea{\end{eqnarray}}
\def\bi{\begin{itemize}}
\def\ei{\end{itemize}}

\def\four{{\textstyle{1\over4}}}

\shorttitle{Moment of inertia testing $e$-capture supernovae}
\shortauthors{Newton et al.}

\begin{document}


\title{Testing the formation scenarios of binary neutron star systems with measurements of the neutron star moment of inertia}

\author{William G. Newton$^{*}$}
\affil{Department of Physics and Astronomy, Texas A\&M University-Commerce, Commerce, TX 75429-3011, USA}
\author{Andrew W. Steiner}
\affil{Department of Physics and Astronomy, University of Tennessee, Knoxville, TN 37996, USA}
\affil{Physics Division, Oak Ridge National Laboratory, Oak Ridge, TN 37831, USA}
\author{Kent Yagi}
\affil{Department of Physics, Princeton University, Princeton, NJ 08544, USA}
\email{$^{*}$william.newton@tamuc.edu}

\begin{abstract}
Two low mass neutron stars, J0737-3039B and the companion to J1756-2251, show strong evidence of being formed from the collapse of an ONeMg core in an electron capture supernova (ECSN) or in an ultra-stripped iron core collapse supernova (FeCCSN). Using three different systematically generated sets of equations of state we map out the relationship between the moment of inertia of J0737-3039A, a candidate for a moment of inertia measurement within a decade, and the binding energy of the two low mass neutron stars. We find this relationship, a less strict variant of the recently discovered I-Love-Q relations, is nevertheless more robust than a previously explored correlation between the binding energy and the slope of the nuclear symmetry energy $L$. We find that, if either J0737-3039B or the J1756-2251 companion were formed in an ECSN, no more than $0.06 M_{\odot}$ could have been lost from the progenitor core, more than four times the mass loss predicted by current supernova modeling. Furthermore, a measurement of the moment of inertia of J0737-3039A to within 10\% accuracy can discriminate between formation scenarios such as ECSN or ultra-stripped FeCCSN and, given current constraints on the predicted core mass loss, potentially rule them out. Using the I-Love-Q relations we find that an Advanced LIGO can potentially measure the neutron star tidal polarizability to equivalent accuracy in a neutron star-neutron star merger at a distance of 200 Mpc, thus obtaining similar constraints on the formation scenarios. Such information on the occurrence of ECSNe is important for population synthesis calculations, especially for estimating the rate of binary neutron star mergers and resulting electromagnetic and gravitational wave signals. Further progress needs to be made modeling the core collapse process that leads to low-mass neutron stars, particularly in making robust predictions for the mass loss from the progenitor core. 
\end{abstract}

\keywords{stars: neutron star  --- pulsars: individual (J0737-3039A/B, J1756-2251) --- stars: evolution --- binaries: close --- dense-matter --- equation of state}

\section{Introduction}

The double pulsar system PSR J0737-3039A/B \citep{Burgay:2003qf,Lyne:2004pd} and the double neutron star system containing PSR J1756-2251 \citep{Ferdman:2014ve} share the following properties, which suggest a specific, shared evolutionary history \citep[see][]{Burgay:2003qf,Lyne:2004pd,Piran:2005bh,Stairs:2006rm,Ferdman:2014ve,Ferdman:2013zl,Iacolina:2016ul}. The second-born neutron stars in both systems have similar gravitational masses: 1.2489$\pm$0.0007~M$_{\odot}$ for PSR J0737-3039B, and 1.230$\pm$0.007~M$_{\odot}$ for the companion neutron star to PSR J1756-2251. The systems' orbital eccentricities (0.088 and 0.18 respectively) are relatively low, as is the alignment angle between the first-born neutron star's spin and the systems' orbital angular momentum ($\lesssim$10$\degree$ and $\lesssim$35$\degree$ respectively).  The systems' transverse velocities are small ($\sim 10$ km/s), and they are both located close to the galactic plane, making it statistically likely that the overall system velocities are small. The first-born neutron stars in both systems have very stable pulse profiles. 

Taken together, these observations suggest that the second-born neutron star in each system was formed in a supernova which delivered a very small kick to the system - i.e. a symmetric explosion which occurred on timescales too short for instabilities to develop, resulted in very little core mass loss, and was sub-energetic \citep{Scheck:2004it,Podsiadlowski:2005cr,Kitaura:2006dp}. A strong candidate progenitor for such an explosion is a low-mass helium star progenitor whose massive hydrogen envelope has been stripped in binary interactions  \citep{Nomoto:1984hc,Podsiadlowski:2005cr,Dewi:2002sp,Ivanova:2003ij}, and which leads to an electron-capture supernova (ECSN) \citep{Podsiadlowski:2005cr,Ferdman:2013zl}, the result of destabilization of an ONeMg core by electron-captures onto magnesium and neon nuclei \citep{Miyaji:1980fk,Nomoto:1987th,Kitaura:2006dp}.

Models of the ECSN progenitor cores suggest the onset of the
electron-capture instability occurs at a unique ONeMg core mass in the
mass range of $1.366 - 1.377$ M$_{\odot}$.
\citep{Miyaji:1980fk,Nomoto:1984hc,Nomoto:1987th,Podsiadlowski:2005cr,Takahashi:2013yu}. Electron
captures cause the core to contract, and O and Ne burning is ignited
in the central regions and propagates outwards in a deflagration front
\citep{Schwab:2015ys}, processing material to nuclear statistical
equilibrium (NSE), where further electron captures and
photdissociation accelerates the collapse
\citep{Miyaji:1980fk,Nomoto:1987th,Takahashi:2013yu}. Whether the core
collapses or the deflagration disrupts the core depends sensitively on
the ignition density \citep{Isern:1991zr,Jones:2016fr}. If the core
does collapse, the explosion proceeds via delayed explosion on short
timescales \citep{Mayle:1988qf,Kitaura:2006dp,Fischer:2010rt}, and 2D
simulations suggest the explosion occurs before significant convection
has had time to develop \citep{Wanajo:2011jk} and hence a symmetric explosion results. This, coupled with the
steep density gradient at the core surface, leads to very little mass
loss from the core; estimates of mass loss include of order $10^{-3}$ M$_{\odot}$
\citep{Podsiadlowski:2005cr}, $10^{-2}$ M$_{\odot}$ \citep{Kitaura:2006dp}, and $1.39
\times 10^{-2}$ M$_{\odot} (1.14 \times 10^{-2}$ M$_{\odot})$ for the 1D
(2D) models of \citet{Wanajo:2009xy,Wanajo:2011jk}. Therefore the ONeMg
progenitor core mass is a good estimate of the baryon mass $M_{\rm B}$
of the resulting neutron star \citep{Podsiadlowski:2005cr}. Indeed, PSR J0737-3039A and the companion to PSR J1756-2251 have gravitational masses consistent with baryon masses $\sim 1.37$ M$_{\odot}$ when their gravitational binding energies are taken into account \citep{Lattimer:1989fk}. Population synthesis calculations incorporating the various binary evolution channels that might lead to production of neutron stars via ECSNe show that J0737-3039B most likely formed in an ECSN, and the companion to PSR J1756-2251 is consistent with such a formation scenario \citep{Andrews:2015az}. Other systems with candidates for ECSNe formation also exist \citep{Keith:2009xe,Chen:2011uo}.

Establishing the existence of an ECSN pathway in stellar evolution is important in a nuclear physics context.
The gravitational binding energy of a neutron star $BE = M_{\rm B} - M_{\rm G}$ is highly sensitive to the neutron star equation of state
(EOS) \citep{Lattimer:1989fk} and it was shown that the constraint on the binding energy of J0737-3039B from the assumption of an ECSN formation scenario gives constraints on EOS \citep{Podsiadlowski:2005cr} and the slope of the nuclear symmetry energy $L = 3 n_0 (dS/dn)_{n_0}$ \citep{Newton:2009qe}, where the nuclear symmetry energy $S(n)$ is the difference between the energies of pure neutron matter (PNM) and symmetric nuclear matter (with a proton fraction of one half; SNM). $n$ is the baryon number density and $n_0 = 0.16$ fm$^{-3}$ nuclear saturation density. Tighter constraints on the nuclear symmetry energy will allow a number of important astrophysical and nuclear quantities to be more accurately determined \citep{Tsang:2012qy,Li:2014fj,Lattimer:2014uq}.

Establishing the existence of an ECSN pathway is also very important in a number of astrophysical contexts. The population of short orbital and pulsar spin period Be/X-ray binaries is suggestive of an ECSN formation for the neutron star \citep{Knigge:2011mz}, and the creation of such systems is signified by an increase in formation efficiency of Be/X ray binaries in starbursts of ages 20-60 Myrs  \citep{Linden:2009ly}. The population of double neutron star systems is also consistent with the majority of second-formed neutron stars in such systems originating in ECSNe \citep{Beniamini:2016ek}. Upper limits on the contribution of ECSNe to all core-collapse supernova have been placed at 20\%-30\% \citep{Poelarends:2008gf,Wanajo:2009xy} taking into account production channels involving single super-asymptotic giant brach (AGB) stars \citep{Siess:2007ve,Poelarends:2008gf,Pumo:2009ul}, and those that take into account binary interactions which may lead to systems such as J0737-3039A/B and J1756-2251 \citep{Podsiadlowski:2004qf,Podsiadlowski:2005cr,Ivanova:2008dn} (with the binary channels likely to dominate). Finding evidence that particular binary systems gave rise to an ECSN will substantially affect such population synthesis calculations, and estimates of the rates of binary neutron star mergers and resulting rate of detectable gravitational wave signals.

Recently, the binary evolutionary pathway that leads to ECSNe has been
subsumed into the broader category of ultra-stripped (US) supernovae
(resulting from a helium star which has been stripped of most of its
envelope by mass transfer onto a companion neutron star), which includes some
iron core-collapse supernovae (FeCCSNe)
\citep{Tauris:2013zz}. The small ejecta mass and resulting symmetric explosion and low kicks
make such SNe additional candidates for the progenitors of PSR
J0737-3039A/B, PSR J1756-2251 and similar systems
\citep{Tauris:2015yu}. Binary evolution simulations suggest that
ultra-stripped FeCCSNe occur in He stars whose metal cores exceed 1.43
M$_{\odot}$ \citep{Tauris:2015yu}. Unlike the progenitor core for
ECSNe, there is no unique Chandrasekhar mass for ultra-stripped
FeCCSNe. Simulations suggest that the lowest mass progenitor metal
cores of ultra-stripped FeCCSNe, $\sim 1.45$ M$_{\odot}$, explode with
core mass loss of $\sim 0.1$ M$_{\odot}$, producing a neutron star with
baryon mass $\sim 1.35$ M$_{\odot}$ \citep{Suwa:2015qf}. More
exploration of the parameter space of progenitors of ultra-stripped
FeCCSNe and their resulting supernovae is necessary, but in this paper
we will take $M_{\rm B}=1.35 M_{\odot}$ as the lowest baryon mass of
neutron stars resulting from a ultra-stripped FeCCSN.
    
Signatures of ultra stripped/ECSNe as distinguished from ordinary FeCCSNe 
may show up in the pre-supernova neutrino signal,
detectable either in specific events \citep{Kato:2015lq}, or on
average from relic neutrinos \citep{Mathews:2014dq}. Ultra stripped/ECSNe are expected
to be faint supernovae \citep[explosion energies around $10^{50}$ ergs][]{Kitaura:2006dp} and may be associated with certain supernova classifications: those that occur in binary systems from progenitors stripped of their hydrogen appear as sub-classes of Type-I SNe such as type Ib/c \citep{Podsiadlowski:2005cr} or type Iax \citep{Moriya:2016rr,Kitaura:2006dp}, whereas those from single super-AGB stars are expected to appear as Type II subclasses such as Type IIn-P \citep{Smith:2013cr,Kitaura:2006dp}. Finally, ECSNe they produce Ni and Fe abundances that can distinguish them from FeCCSNe \citep{Wanajo:2009xy,Smith:2013cr}. \emph{In this paper we will demonstrate an additional method to distinguish systems which hosted an ECSN or ultra-stripped FeCCSN from those that didn't.}   

Strong constraints on the equation of state can be obtained from a measurement of a neutron star's moment of inertia \citep{Ravenhall:1994zl,Lattimer:2001zz,Morrison:2004zz,Bejger:2005fv,Worley:2008zz,Fattoyev:2010yg,Steiner:2015rw,Raithel:2016lr} or its tidal polarizability $\lambda$ \citep{Postnikov:2010qy,Hinderer:2010uq,Pannarale:2011fj,Fattoyev:2013ij}. It has been noted that a simultaneous measurement of the moment of inertia of PSR J0737-3039A and the binding energy of PSR J0737-3039B, would lead to a stronger constraint on the equation of state \citep{Morrison:2004zz} compared to constraints from individual measurements. The measurement of the spin-orbit contribution to the precession of the periastron of the double pulsar system J0737-3039 could be accurate enough to infer the moment of inertia of pulsar A to within 10\% in the next few years \citep{Damour:1988jk,Lattimer:2005vn,Kramer:2009lr}, although this depends on the accuracy of a number of Post-Keplerian parameters of the system \citep{Iorio:2009rt}. The most recent estimates predict that such an accurate measurement will be possible in the next decade, taking into account next generation radio telescopes such as SKA \citep{Kehl:2016ys}. It is estimated that Advanced LIGO-Virgo should obtain constraints on the tidal polarizability of a neutron star from the gravitational waveform of a binary neutron star-neutron star merger or neutron star-black hole merger \citep{Read:2009yq,Pannarale:2011fj,Hotokezaka:2016kx}. Particularly, the dimensionless tidal polarizability $\lambda/M_{\rm G}^5$ can be measured to a 1-$\sigma$ accuracy of 100 by an Advanced LIGO detection of  a merger at a distance of 200 Mpc \citep{Hotokezaka:2016kx}; this is equivalent to a measurement of the dimensionless moment of inertia of $I/M_{\rm G}^3$ of less than 10\% accuracy independent of the stiffness of the EOS \citep{Yagi:2016lr}.

Recently, universal (EOS-independent) relations between the moment of inertia and a number of other global neutron star parameters such as the mass quadrupole moment and tidal polarizability have been found: the so-called I-Love-Q relations \citep{Yagi:2013lr,Yagi:2013qv}; and see \citet{Yagi:2016lr} for a recent review on other universal relations. Particularly relevant for this work, is the relation between moment of inertia and binding energy \citep{Steiner:2016zr}, which, although not displaying  quite the same universality as the I-Love-Q relations, nevertheless enables a moment of inertia measurement to place constraints on the binding energy of a neutron star. In this paper we illustrate how these relations enables a measurement of the moment of inertia of PSR J0737-3039A to potentially constrain the ultra-stripped/ECSN scenario of PSR J0737-3039B and the companion to PSR J1756-2251, by examining in detail the relation between the moment of inertia of a $1.338$ M$_{\odot}$ neutron star (corresponding to the mass of PSR J0737-3039A) and the binding energy of a neutron star with the masses of PSR J0737-3039B and the companion to PSR J1756-2251.

The organization of the rest of the paper is as follows. In section~(\ref{sec:EOS}) we describe the 3 models we employ to systematically explore the parameter space of the equation of state at high and low densities. In section~(\ref{sec:results}) we describe the resulting correlations between the binding energy and moment of inertia, and between the binding energy and slope of the symmetry energy, and discuss the implications for testing the ECSN and ultra-stripped FeCCSN scenarios. In section~(\ref{sec:con}) we present our conclusions.

\section{Equations of state}\label{sec:EOS}

The proton fraction in the neutron
star outer core is sufficiently small that the EOS is well
approximated by the pressure of pure neutron matter (PNM). There has
been significant recent
progress~\citep{Gandolfi:2012lr,Hebeler:2013zl,Gezerlis:2013lr} on
computing the EOS of PNM from realistic nuclear forces, using the
quantum Monte Carlo method and chiral effective theory interactions in
many-body perturbation theory. 
The first two EOSs both use the
parameterization of the results of the quantum Monte Carlo model
from~\cite{Gandolfi:2012lr} given in~\cite{Steiner:2012kq}, and we
refer to this model as ``GCR''. The limits on the parameters of that
model, $12.5~\mathrm{MeV}<a<13.5~\mathrm{MeV}$ and $0.47<\alpha<0.53$
are as used in~\citep{Steiner:2015rw}. These two parameters
principally parameterize the two-nucleon part of the interaction.
Also, as in~\cite{Steiner:2015rw}, we reparametrize $b$ and $\beta$,
parameters which control the three-nucleon interaction, in terms of the magnitude and slope of the symmetry energy $S$
and $L$. Particularly, we limit $L$ to be between 30 and 70 MeV which
covers the range of $L$
from~\cite{Gandolfi:2012lr,Gandolfi:2014xy,Steiner:2015rw}.


The first two EOSs are the same
near the saturation density but differ at high-density. We attach the
GCR results either to a set of three piecewise polytropes referred to
as ``GCR+Model A'' in
~\citep{Steiner:2013lq,Steiner:2015rw} or to a set of four
line segments in the $(\varepsilon,P)$ plane, ``GCR+Model
C''~\citep{Steiner:2013lq,Steiner:2015rw}. This latter model is useful
because it provides an alternative model which tends to favor stronger
phase transitions in the core.

The third set of EOSs use the Skyrme energy-density functional to construct the EOS near saturation density. In previous work \citep{Fattoyev:2012hc}, we developed families of Skyrmes by taking a baseline parameterization and refitting the two purely isovector model parameters $x_0$ and $x_3$ to the latest results of PNM calculations \citep{Gezerlis:2010fu,Hebeler:2010kb,Gandolfi:2012lr,Gandolfi:2015nr,Lynn:2016sf}. The resulting re-fit Skyrme models follow a tight correlation  $S = 0.1 L + 26.4$ MeV between the slope and magnitude of the symmetry energy, and we explore a wide range 20$<L<$120 MeV. The fact that only the two purely isovector parameters are adjusted means that such adjustments leave SNM properties unchanged \citep{Chen:2009it}.
These Skyrme EOSs are used up to 1.5 $n_0$, and replaced at higher with two piecewise polytropes, with a total of 2 free parameters (after the transition to the first polytope is fixed at 1.5 $n_0$ by the pressure of the Skyrme EOS there). This completes the third set of EOSs which we label ``Skyrme+Poly''.

There are two main differences between the Skyrme+Poly EOSs and
GCR+Models A,C. Firstly, the Skyrme EOSs explore a wider range of possible values of the slope of the symmetry energy $L$ than GCR, although it is important to note that the highest and lowest values of $L$ give poor fits to the current results of microscopic pure neutron matter calculations. Secondly, The Skyrme EOSs have a well defined symmetry energy curvature $K_{\rm sym}$ which correlates linearly with $S$ and $L$ after the two isovector Skyrme parameters have been fit to PNM. A soft $L$ leads to a soft EOS up to at least 1.5 $n_0$ where the first polytrope is attached. In contrast, EOSs GCR+Models A,C adjoin the first of their polytr1opes or line segments at saturation density, thus decoupling $K_{\rm sym}$ from $L$ and $S$.

The parameter space for GCR+Model A and C EOSs is explored by
performing a Markov chain Monte Carlo simulation as first outlined in
~\citep{Steiner:2010bh} and implemented in
\citep{Steiner:2014fk,Steiner:2014qy}
using uniform priors in the model parameters and with the only
astrophysical constraints on the neutron star (NS) maximum mass,
$M_{\rm max}\geq2.0M_{\odot}$, and
the constraint that matter is stable and causal. To obtain our final results we
choose the smallest range in the EOS parameter space which encloses all of the EOS models, as
done in~\citep{Steiner:2013lq}. 

The parameter space of the Skyrme+Poly EOSs is explored as follows. We
select 7 baseline Skyrmes which lie in the ranges of binding energy,
saturation density and incompressibility of symmetric nuclear matter
as the GCR model, and which lie along the $S-L$ correlation created by
the re-fit to PNM calculations at values of $L$ equal to 20, 40, 50, 60, 70, 80 and 120 MeV. Then, for each value of $L$ we (1) adjust the two free polytropic parameters to obtain a particular maximum masses starting at 2.0 $M_{\odot}$, up to the mass at which causality is violated in the center of the star; (2) at this fixed value of $L$ and $M_{\rm max}$, we adjust the two parameters to obtain the maximum and minimum moments of inertia $I$ of a 1.4$M_{\odot}$ NS without the EOS violating causality at any point below the central density of the maximum mass model. For the Skyrme models, we are effectively using uniform priors in the quantities that parameterize the family of Skyrme+Poly equations of state: $L$, $M_{\rm max}$ and $I$.

\section{Results and Discussion} \label{sec:results}

%
%
\begin{figure}
\begin{center}
\includegraphics[width=6cm,height=6cm]{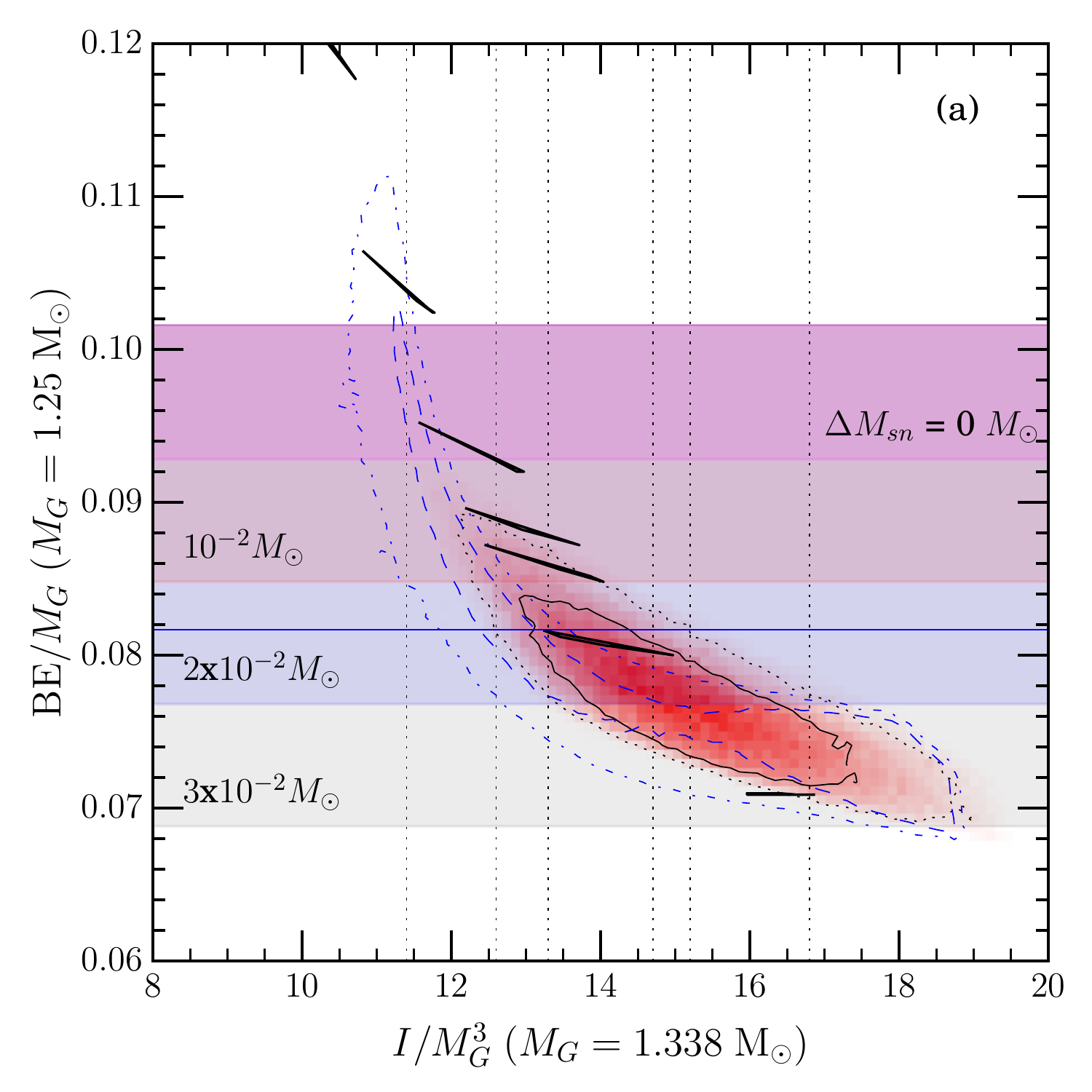}\includegraphics[width=6cm,height=6cm]{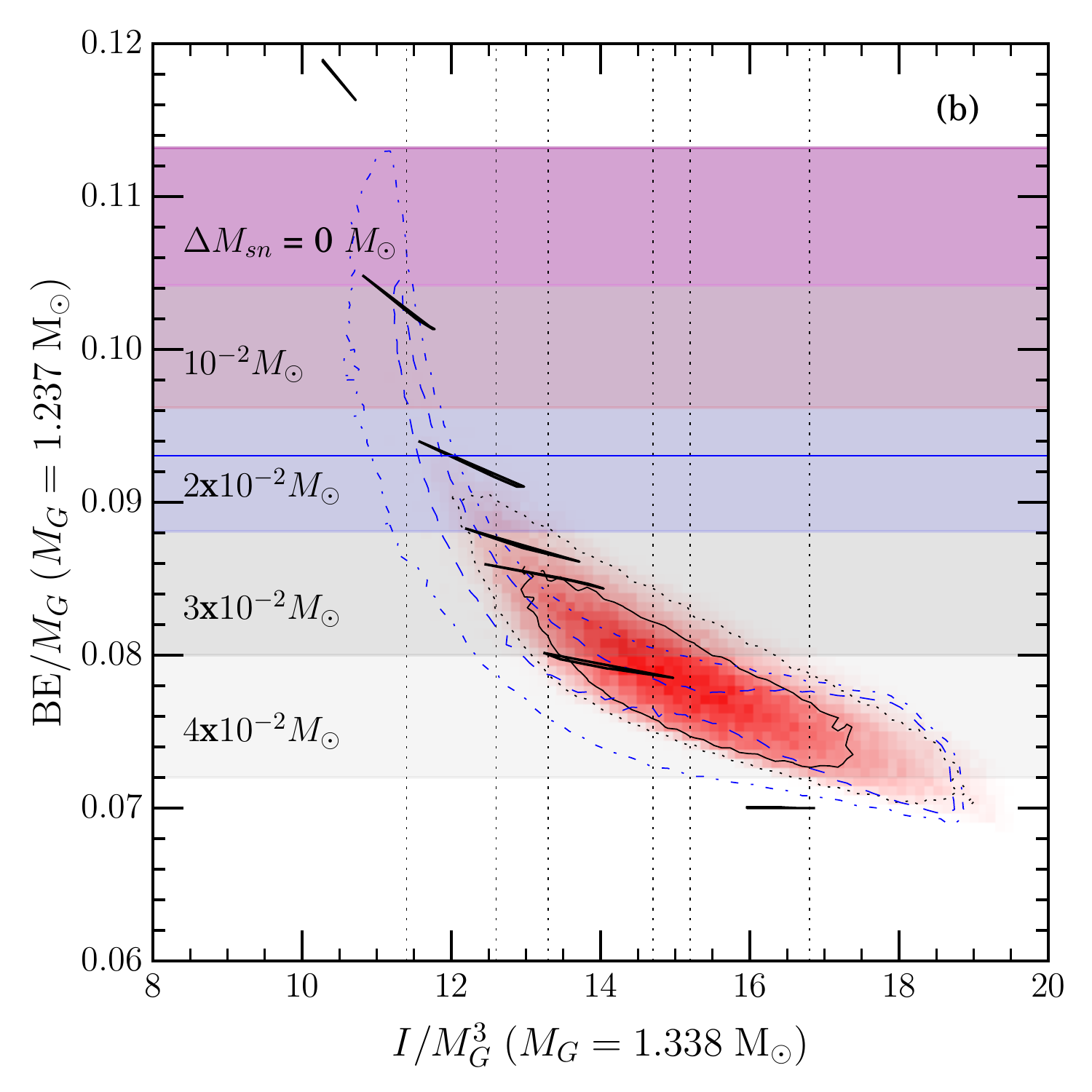}\includegraphics[width=6cm,height=6cm]{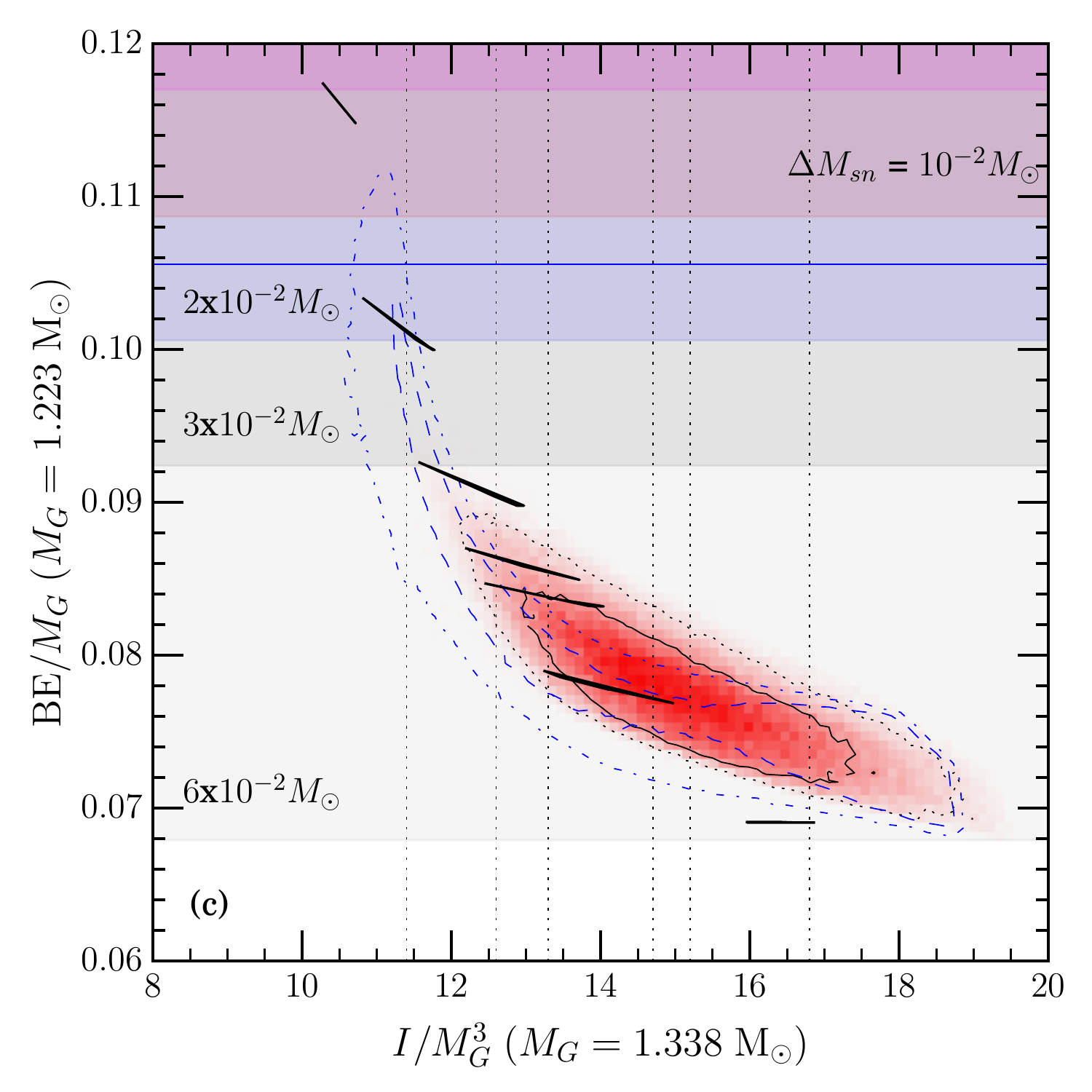}
\caption{
\label{fig:1}
(Color online) Dimensionless binding energy $BE/M_{\rm G}$ versus dimensionless moment of inertia $I/M_{\rm G}^3$. The binding energy is that of a 1.25$M_{\odot}$ neutron star (mass of J0737-3039B; a), 1.237$M_{\odot}$ neutron star (upper limit on mass of the companion to J1756-2251; b) and a 1.223$M_{\odot}$ neutron star (lower limit on mass of the companion to J1756-2251; c). The moment of inertia is that of a 1.338$M_{\odot}$ neutron star (mass of J0737-3039A, a candidate for a moment of inertia measurement to within 10\% in the next $\sim$10 years). On each graph we plot the set of systematically generated Skyrme EOSs as the short black bands. Each individual band represents a certain value of the slope of the symmetry energy $L$ from 20 MeV (highest binding energy) to 120 MeV (lowest binding energy). Each band is generated by systematically varying the stiffness of the two high-density polytopes to (i) obtain maximum masses of 2.0 to 2.6$M_{\odot}$ and then to (ii) for each maximum mass, obtain the highest and lowest possible values of moment of inertia consistent with causality. In addition, we plot the probability density distribution in red for the Model A EOSs together with the Model A 68\% confidence limit (black solid contour) and 95\% confidence limit (black dotted contour) and the Model C 68\% confidence limit (blue dashed contour) and 95\% confidence limit (blue dash-dotted contour). Model C allows for stronger phase transitions. Constraints on the binding energy assuming an ECSN origin for the respective pulsars are given by the colored bands with different amounts of mass loss during the collapse $\Delta M_{\rm sn}$. The width of each band represents the uncertainty in the pre-collapse progenitor core mass. The lower limit on the dimensionless binding energy from current modeling of the progenitor core plus ECSN is given by the blue horizontal line. Three hypothetical measurements of the moment of inertia of J0737-3039A to 10\% accuracy are given by the 3 sets of vertical dotted lines: $I/M_{\rm G}^3 = 12 \pm 0.6, 14 \pm 0.7$ and $16 \pm 0.8$. Given that current modeling predicts mass loss of up to $\sim 1.4 \times 10^{-2} M_{\odot}$, a measurement of the moment of inertia to within 10\% can potentially rule out an electron capture supernova origin of the pulsars if the measured moment of inertia is sufficiently large.}
\end{center}
\end{figure}

The binding energy is calculated as $BE = M_{\rm B} - M_{\rm G}$ where $M_{\rm G}$ is the gravitational mass and $M_{\rm B}$ is the baryon mass, given by $M_{\rm B} = A m_{\rm B}$ where $A$ is the baryon number of the star, and $m_{\rm B}$ is the mass per baryon. For the ONeMg progenitor core of an ECSN, $m_{\rm B} \approx 931.5$ MeV; for an Fe core, $m_{\rm B} = 930.4$ MeV. The moment of inertia is calculated in the slow rotation approximation \citep{Hartle:1967fk,Hartle:1968qf}, and the dimensionless tidal polarizability $\lambda/M_{\rm G}^5$ is calculated from the dimensionless moment of inertia $I/M_{\rm G}^3$ using the universal relations of \citet{Yagi:2016lr}.

%
%
\begin{figure}
\begin{center}
\includegraphics[width=6cm,height=6cm]{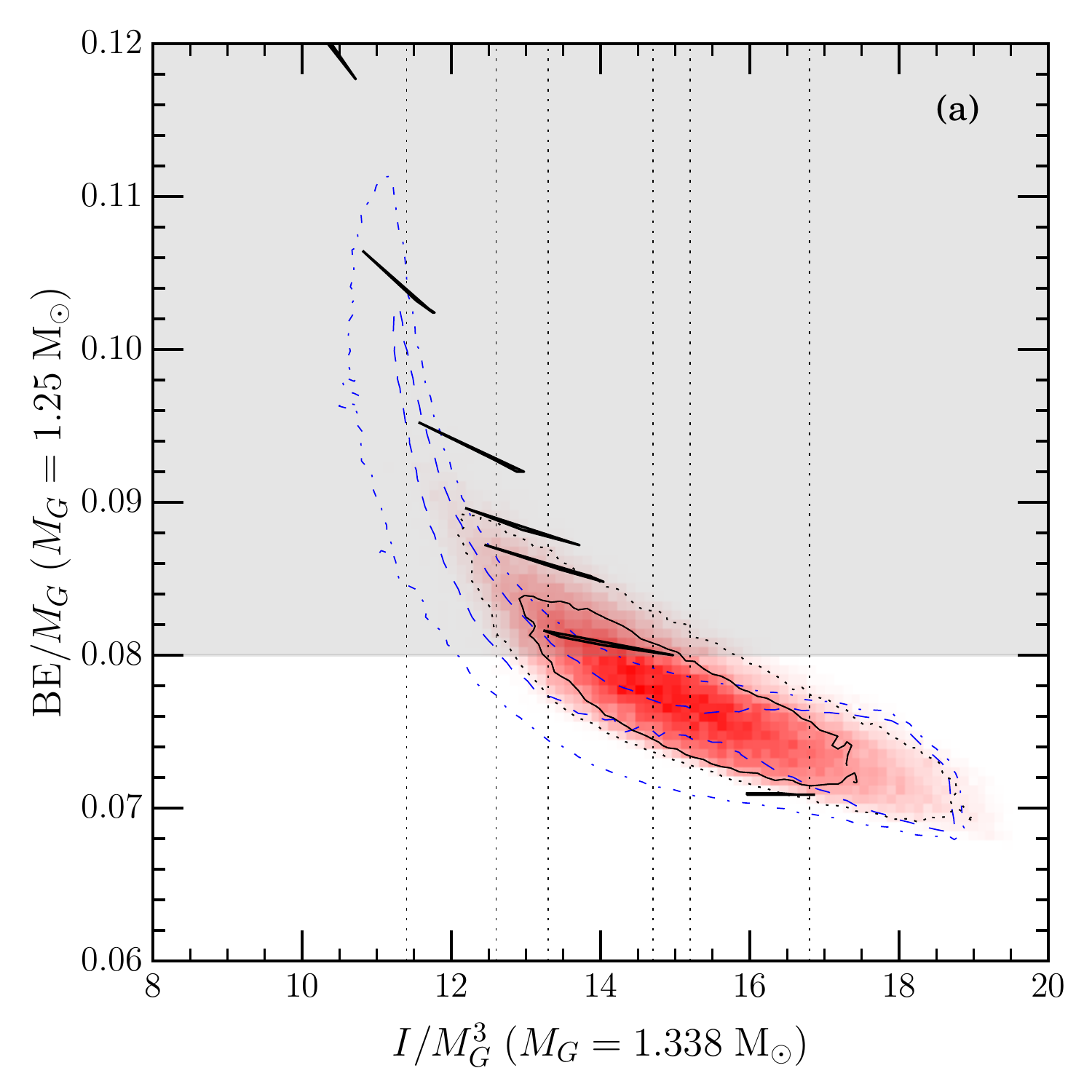}\includegraphics[width=6cm,height=6cm]{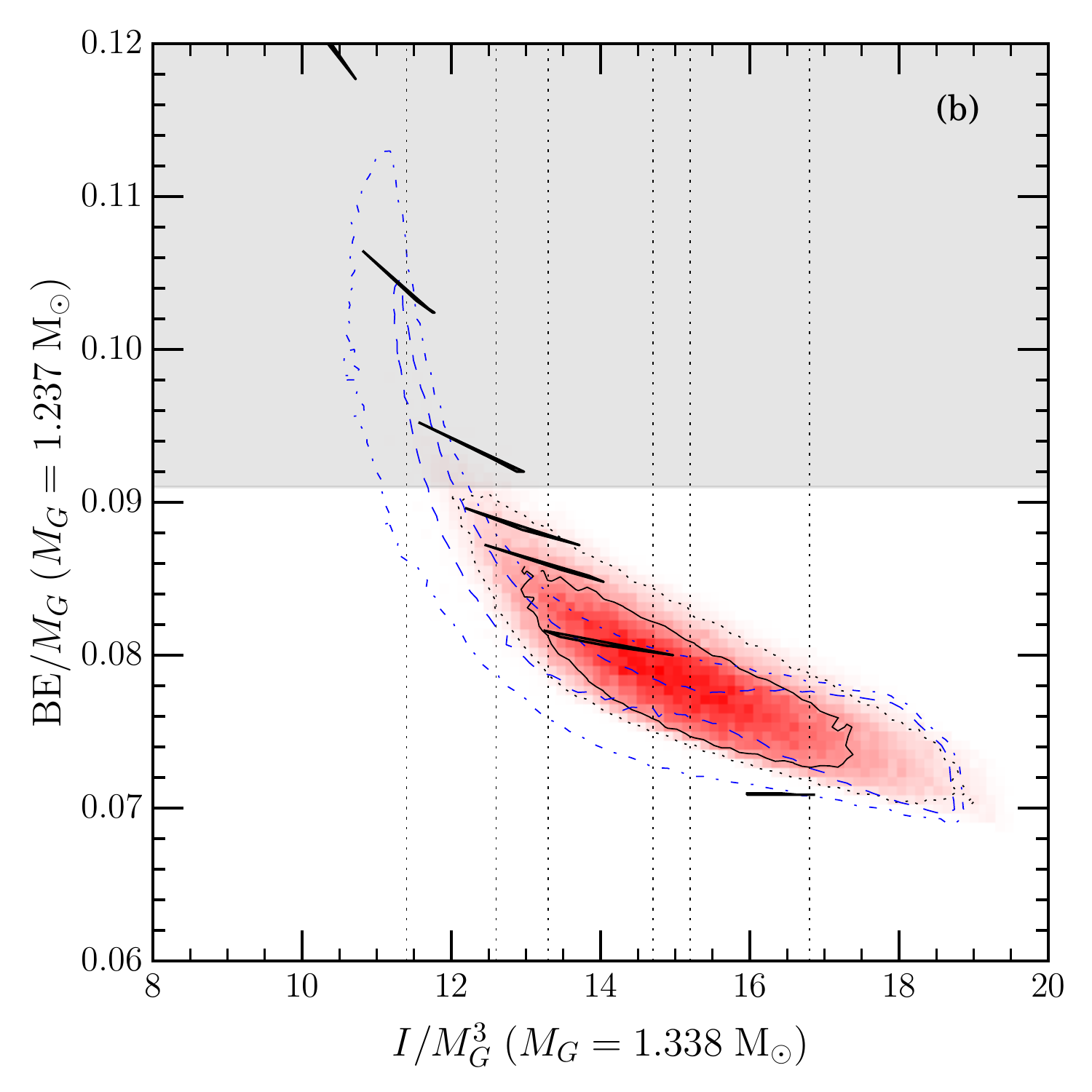}\includegraphics[width=6cm,height=6cm]{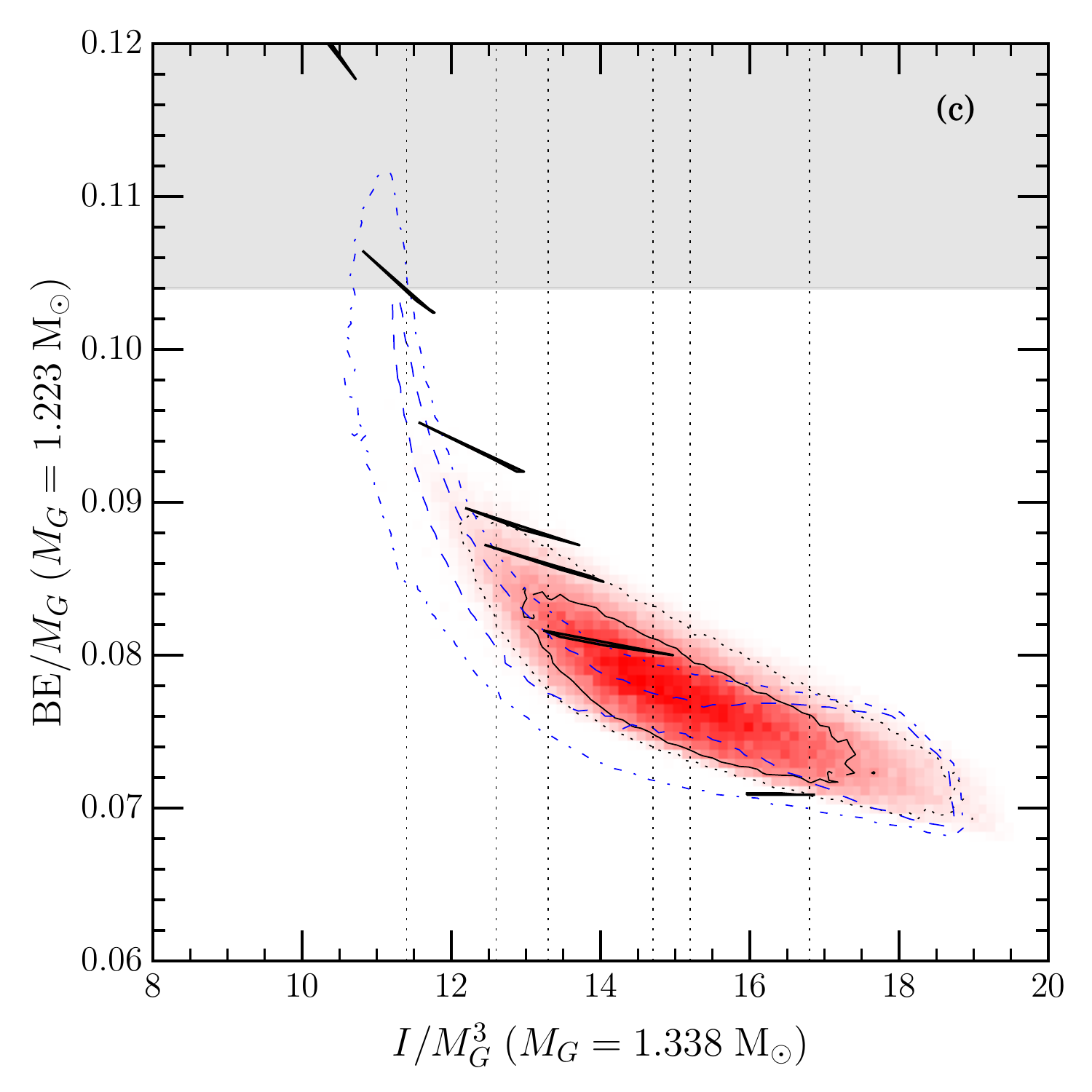}
\caption{
\label{fig:2}
Same as Fig.~\ref{fig:1}, but now the shaded band indicates the lower limit on the binding energy of a neutron star of the respective mass assuming it is born in an ultra-stripped iron core collapse supernova, as inferred from supernova modeling. A 10\% measurement of the moment of inertia of J0737-3039A can rule out such a formation scenario, assuming current modeling of the supernova is robust and complete if the measured moment of inertia is sufficiently large.}
\end{center}
\end{figure}

In Figs.~\ref{fig:1} and~\ref{fig:2}, we plot the relations between the dimensionless binding energy $BE/M_{\rm G}$ and the dimensionless moment of inertia $I/M_{\rm G}^3$. The results from the Skyrme+Poly models are given by the short black bands, with each line representing a different value of $L=20,40,50,60,70,80,120$ MeV with the lower $L$ giving the higher binding energy. Each band is the locus of points representing the equations of state with masses above $2.0 M_{\odot}$ and, for each maximum mass, equations of state with moments of inertia between the minimum and maximum allowed by causality. The GCR+Model A EOSs are represented by the red density distribution, solid 68\% confidence contour and dotted 95\% confidence contours, and GCR+Model C represented by the dashed 68\% confidence contours and dash-dotted 95\% confidence contours. The moment of inertia is that of a neutron star of the mass of J0737-3039A, 1.338 $M_{\odot}$, and we present results for binding energies of neutron stars with the mass of J0737-3039B, 1.25 $M_{\odot}$, and the upper and lower limits to the mass of the companion to J1756-2251, 1.237 and 1.223 $M_{\odot}$, in the left, center and right columns respectively. In Fig.~\ref{fig:1}, the colored bands represent the constraint on the binding energy from the assumption of an ECSN creation scenario for the pulsars under different assumptions of the mass lost from the core during the explosion $\Delta M_{\rm sn}$. The $\Delta M_{\rm sn}$=0 M$_{\odot}$ band shows the binding energy range inferred from the progenitor mass range of $1.366 - 1.377$ M$_{\odot}$ \citep{Nomoto:1984hc,Podsiadlowski:2005cr,Takahashi:2013yu}. Current modeling suggests $\Delta M_{\rm sn} \lesssim 1.5 \times 10^{-2}$ M$_{\odot}$.
In Fig.~\ref{fig:2} we show the same relations, now with the lower limit on the binding energy from the limited constraints from modeling of ultra-stripped FeCCSN, shown as the grey bands. 

We see that the general trend, as expected, is that higher binding energy (corresponding to more compact neutron stars) corresponds to smaller moment of inertia. The distribution of Skyrme+Poly models of different $L$ values are distributed relatively uniformly from high binding energy, small moment of inertia (small $L$) down to small binding energy, large moment of inertia (large $L$). For GCR+model A, the distribution of model parameters is also relatively uniform, and the 95\% confidence contour overlaps with the higher $L$ Skyrme+Poly models ($L = 60 - 120$ MeV). The GCR+Model C allows for stronger phase transitions which generate more models with large binding energy and small moment of inertia, reaching the part of the diagram populated by the low $L$ Skyrme+Poly models, although the highest density of models is still found in the small binding energy, high moment of inertia part of the diagram.

%
%
\begin{figure}
\begin{center}
\includegraphics[width=8cm,height=7cm]{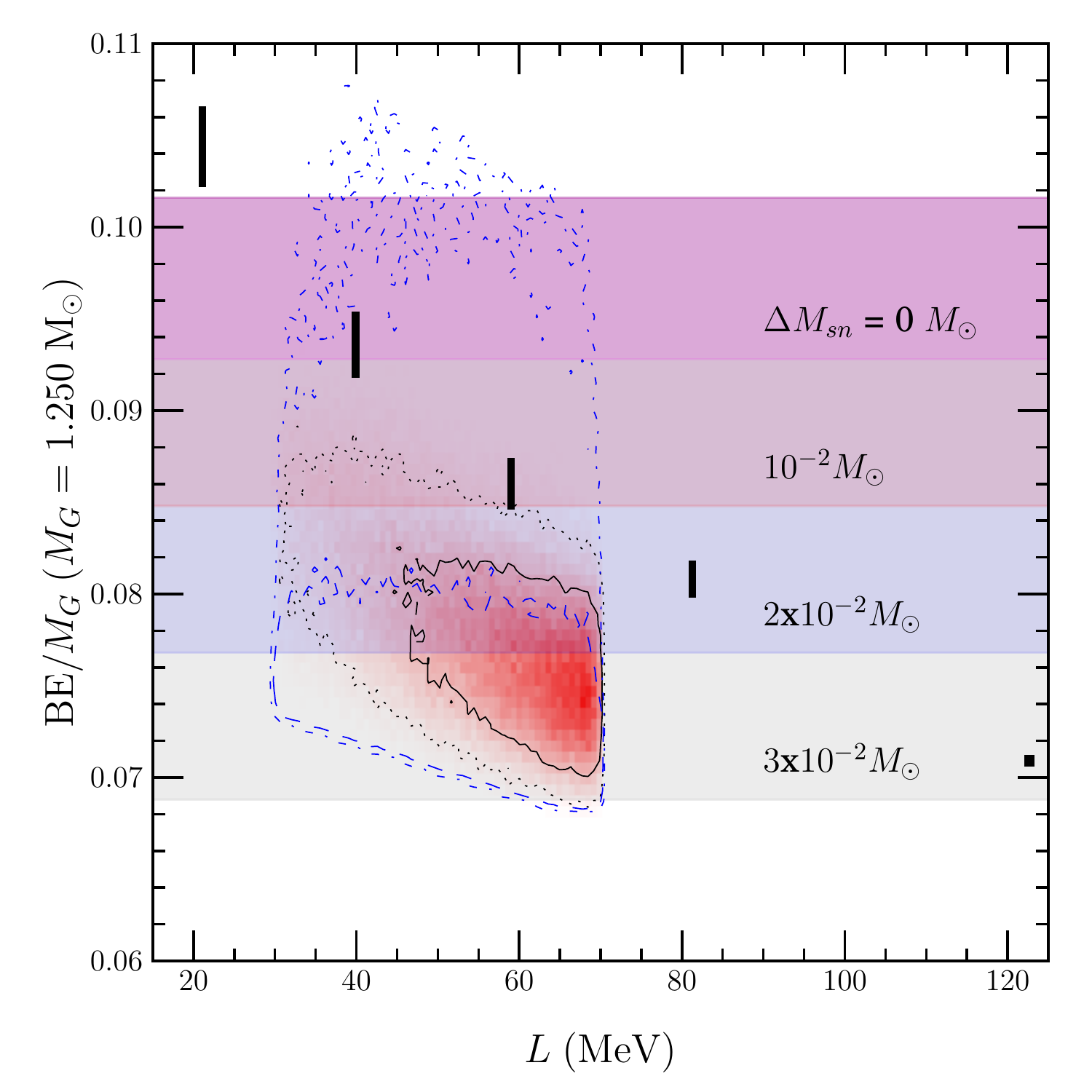}
\caption{
\label{fig:3}
Binding energy of a 1.25 $M_{\odot}$ neutron star versus the slope of the nuclear symmetry energy $L$. As in Figs.~\ref{fig:1} and~\ref{fig:2}, the short black lines are the results of the Skyrme EOSs while the red density distribution and black solid (68\% confidence) and dotted (95\% confidence) contours are the results of Model A, and the dashed (68\% confidence) and dash-dotted (95\% confidence) blue contours are the results of Model C. Constraints on the binding energy assuming an electron capture supernova origin for the respective pulsars are given by the colored bands as in Fig.~\ref{fig:1}.}
\end{center}
\end{figure}

Both GCR+models A and C have symmetry energy slopes between $L=30$ MeV and $L=70$ MeV, yet most models overlap with the high $L$ Skyrme+Poly EOSs $L>70$MeV. This is illustrated by Fig.~\ref{fig:3}, which shows the relationship between dimensionless binding energy $BE/M_{\rm G}$ and the slope of the symmetry energy $L$, with the results from the different EOS models displayed in the same way as Figs.~\ref{fig:1} and~\ref{fig:2}. The binding energies of GCR+models A and C are systematically smaller than those of the Skyrme+Poly models at a given $L$. This is because GCR+models A and C allow for a weak or strong transition at saturation density, where the first polytrope or line segment is attached to the GCR EOS. For the Skyrme+Poly EOSs, a small value of $L$ at saturation density (soft EOS just below saturation density) will lead to a soft EOS up to 1.5 $n_{\rm s}$, and a larger binding energy. However, for GCR+models A and C, a small $L$ at saturation density is compensated in many models by a transition to a stiffer EOS between $n_{\rm s}$ and 1.5$n_{\rm s}$ and a correspondingly smaller binding energy. Since the requirement of $M_{\rm max} > 2.0$ M$_{\odot}$ tends to select stiffer EOSs in this region, there are many more models in the model space with these characteristics, and this is reflected in the probability density plots in Figs.~\ref{fig:1} and~\ref{fig:2}. Therefore, the low $L$ GCR+models A and C with a stiff first polytrope will match the binding energy of the higher $L$ Skyrme+Poly EOSs. The moment of inertia is sensitive most to the EOS just above saturation density, and so these models will also give similar moments of inertia. Thus the correlation between $BE/M_{\rm G}$ and $I/M_{\rm G}^3$ is preserved. 

Furthermore, in GCR+Model A, models cluster at higher $L$ (60-70 MeV) and smaller binding energy (there are more models with a stiffer symmetry energy that satisfy $M_{\rm max} > 2.0$ M$_{\odot}$). For GCR+Model C, the possibility of a strong phase transition near saturation density more strongly decouples the high and low density EOSs, resulting in a binding energy that is largely independent of $L$.

The correlation between $I$ and $BE$ is more robust than the correlation between $L$ and $BE$ explored in \citet{Newton:2009qe}. However, the prediction of \citet{Newton:2009qe} that $L<$70 MeV is consistent with the ECSN scenario assuming a mass loss of $< 0.015$ M$_{\odot}$.

Some general conclusions can be made independent of a measurement of moment of inertia. Let us assume the neutron stars PSR J0737-3039B and the companion to PSR J1756-2251 were formed in an ECSN. From Fig.~\ref{fig:1}, they (and any neutron star with a gravitational mass of $\approx 1.25 M_\odot$) have a minimum $BE/M_{\rm G}$ of $\approx 0.07$ or equivalently a minimum binding energy of $\approx 0.085 M_{\odot}$. For PSR J0737-3039B (Fig.~\ref{fig:1}a,b) this corresponds to a maximum of $\approx 0.03$ M$_{\odot}$ mass loss from the progenitor core. Taking the upper limit on the mass of the companion to PSR J1756-2251 (Fig.~\ref{fig:1}c,d), the maximum progenitor core mass loss is $\approx 0.04$ M$_{\odot}$, and taking the lower limit (Fig.~\ref{fig:1}e,f) the maximum progenitor core mass loss is $\approx 0.06$ M$_{\odot}$. These upper limits on the mass loss from these systems are 2-4 times larger than  the current estimates of mass loss from supernova modeling of up to $\approx 0.015$ M$_{\odot}$.

Let us now examine the consequences of specific measurement of the moment of inertia to within 10\%. Corresponding ranges for the dimensionless tidal polarizability are given according to the I-Love-Q relations.

\begin{itemize}

\item[(i)] \emph{$I/M_{\rm G}^3=16 \pm 0.8$ (a relatively stiff EOS on average; corresponding to a tidal polarizability of $\lambda/M_{\rm G}^5 \approx 850-1220$)}: in this case, the binding energy of the stars is constrained to be $\approx0.085-0.1$M$_{\odot}$. Under the ECSN scenario, this constrains PSR J0737-3039B (Fig.~\ref{fig:1}a,b) to have been created by the collapse of a core from which $\gtrsim 0.015$M$_{\odot}$ has been lost; this is just inconsistent with current SN models, and would require updated estimates or it would rule out the ECSN scenario. Taking the upper limit of the mass of the companion to PSR J1756-2251 (Fig.~\ref{fig:1}c,d), the mass loss is constrained to be $\gtrsim 0.025$M$_{\odot}$. Taking the lower limit of the mass of the companion to PSR J1756-2251 (Fig.~\ref{fig:1}e,f), the mass loss is constrained to be $\gtrsim 0.045$M$_{\odot}$, over three times the predicted mass loss from the progenitor core in the ECSN scenario. In all three cases, more core mass loss than is currently predicted for a ECSN explosion would be needed, or the ECSN scenario is inconsistent with this measurement of the moment of inertia. From Fig.~\ref{fig:2} we see that such a measurement would be marginally consistent with the production of J0737-3039B in an ultra-stripped FeCCSN, but would rule out production of the neutron star companion to J1756-2251 in this way. An Advanced LIGO detection of a NS-NS merger at a distance of 200 Mpc should be able to measure the tidal polarizability to equivalent accuracy at the 2-$\sigma$ level \citep{Hotokezaka:2016kx}.

\item[(ii)] \emph{$I/M_{\rm G}^3=14 \pm 0.7$ (an EOS of intermediate stiffness on average; $\lambda/M_{\rm G}^5 \approx 520-760$)}: in this case, the binding energy is constrained to be $\approx$0.09-0.11M$_{\odot}$. Under the ECSN scenario, this constrains PSR J0737-3039B (Fig.~\ref{fig:1}a,b) to have been created by the collapse of a core from which $\gtrsim 0.005$M$_{\odot}$ has been lost, which is consistent with current SN models. Taking the upper limit of the mass of the companion to PSR J1756-2251 (Fig.~\ref{fig:1}c,d), the mass loss is constrained to be $\gtrsim 0.025$M$_{\odot}$. Taking the lower limit of the mass of the companion to PSR J1756-2251 (Fig.~\ref{fig:1}e,f), the mass loss is constrained to be $\gtrsim 0.04$M$_{\odot}$. Therefore, more mass would need to be lost from the progenitor core of PSR J1756-2251 than is currently predicted during the ECSN explosion, or the ECSN scenario is inconsistent with this measurement of the moment of inertia. From Fig.~\ref{fig:2} we see that such a measurement would be marginally consistent with the production of J0737-3039B in an ultra-stripped FeCCSN, but would rule out production of the neutron star companion to J1756-2251 in this way.  An Advanced LIGO detection of a NS-NS merger at a distance of 200 Mpc should be able to measure the tidal polarizability to equivalent accuracy at the 2-$\sigma$ level \citep{Hotokezaka:2016kx}.

\item[(iii)] \emph{$I/M_{\rm G}^3=12 \pm 0.6$ (a relatively soft EOS on average; $\lambda/M_{\rm G}^5 \approx 290-430$)}: in this case, the binding energy is constrained to be $\approx$0.105-0.14M$_{\odot}$. Under the ECSN scenario, this is consistent with between $0$ and $0.02$M$_{\odot}$ being lost from the progenitor core of PSR J0737-3039B (Fig.~\ref{fig:1}a,b), and therefore is consistent with current SN modeling.  Taking the upper limit of the mass of the companion to PSR J1756-2251 (Fig.~\ref{fig:1}c,d), the measurement is consistent with between $0$ and $0.03$M$_{\odot}$ being lost from the progenitor core, which is again consistent with current SN modeling. Taking the lower limit of the mass of the companion to J1756-2251 (Fig.~\ref{fig:1}e,f), the measurement is consistent with between $0.015$ and $0.05$M$_{\odot}$ being lost from the progenitor core, marginally inconsistent with current supernova modeling. Thus this measurement of the moment of inertia would be consistent with J0737-3039B being formed in an ECSN, and consistent with the companion to J1756-2251 being formed this way provided its mass does not fall very close to the current lower bound. From Fig.~\ref{fig:2} we see that such a measurement would be consistent with the production of J0737-3039B in an ultra-stripped FeCCSN, and marginally consistent with the production of the neutron star companion to J1756-2251 in this way provided the mass of the pulsar lie in the higher half of the current uncertainty range.  An Advanced LIGO detection of a NS-NS merger at a distance of 200 Mpc should be able to measure the tidal polarizability to equivalent accuracy at the 1-$\sigma$ level \citep{Hotokezaka:2016kx}.

A small value for $I/M_{\rm G}^3 \lesssim 12$ favors a strong phase transition at relatively low densities (model C) or very soft symmetry energy $L\lesssim 40$ MeV.

\end{itemize}

\noindent These results demonstrate the feasibility of ruling out formation scenarios based on a future moment of inertia measurement of a neutron star. They do rely, however, on the robust modeling of the progenitor core mass at the time of collapse, and the supernova itself, particularly the mass lost from the core during the explosion.

\section{Conclusions} \label{sec:con}

Evidence for or against the electron-capture supernova (ECSN) or ultra-stripped iron-core collapse supernova (FeCCSN) formation scenarios of neutron stars in double neutron star systems has an important impact on population synthesis calculations and resulting estimates of rate of merger of double neutron star systems and resulting electromagnetic and gravitational wave signals. The ECSN formation scenario has implications for the number of Be-X-ray binaries and the interpretation of galactic starburst. Strong circumstantial evidence exists that PSR J0737-3039B and companion to J1756-2251 were formed in an ECSN or an ultra-stipped FeCCSN. However, recent modeling of the progenitor ONeMg cores of ECSN demonstrate the sensitivity of this scenario to electron-capture rates, and demonstrate the possibility that such cores will be entirely disrupted before they collapse to form neutron stars. In this paper we have demonstrated that a future measurement of the moment of inertia of a neutron star to within 10\% can potentially provide strong evidence against such scenarios, or constrain certain details such as the mass lost from the collapsing core during such supernova explosions.

A previous study \cite{Newton:2009qe} using a wide variety of existing neutron star equations of state found a correlation between the slope of the nuclear symmetry energy $L$ and the binding energy of neutron stars, and used it to determine that $L\leq70$ MeV if pulsar J0737-3039B was formed in an ECSN with up to $\sim 10^{-2}$ $M_\odot$ mass loss from the progenitor core during collapse. The EOSs used had maximum masses of 1.44$M_{\odot}$ and above, reflecting the highest accurately measured pulsar mass of the time. In this work we have used systematically constructed equations of state to explore a wider range of EOS parameter space, constrained only by causality and the requirement that $M_{\rm max} > 2.0 M_{\odot}$. Two different approaches are taken to construct the equations of state. The first uses a simple parameterization of the EOS up to saturation density determined directly from the results of microscopic pure neutron matter calculations with $30 < L < 70$MeV, supplemented above saturation density with either 3 polytropes (Model A) or, to accommodate the possibility of a strong phase transition, 4 line segments (Model C). The second uses Skyrme EOSs up to 1.5 times saturation density, which are fit to the results of PNM calculations at a given value of $L$ which is allowed to vary over the wider range $20 < L < 120$MeV, and supplemented with 2 polytropes above 1.5 $n_{\rm s}$ which are adjusted to systematically give maximum masses above 2.0$M_{\odot}$ and minimum and maximum moments of inertia.

The most important point is that although the correlation between $L$ and $BE$ first considered in \citet{Newton:2009qe} is no longer strong now we are systematically exploring a wider range of EOS model space, the correlation between $BE/M_{\rm G}$ and $I/M_{\rm G}^3$, while not as strong as some of the other universal relations recently found, is still robust. The result from \cite{Newton:2009qe}, that symmetry energy slope values of $L>70$ MeV favor mass loss greater than that predicted by current ECSN modeling, is still consistent with the results obtained here. 

We find that if J0737-3039B was formed in an ECSN, no more than $0.03 M_{\odot}$ could have been lost from the progenitor core, (more than twice the mass loss predicted by current supernova modeling) and if the companion to J1756-2251 was formed that way, no more than $0.06 M_{\odot}$ could have been lost from the progenitor core during explosion. We demonstrate that a 10\% measurement of the moment of inertia is sufficient to provide evidence for or against the ECSN or ultra-stripped FeCCSN in these two systems, and any similar systems that might be discovered. If such a measurement favors the stiffer EOS models, it can be sufficient to rule out production of the neutron stars in the two formation scenarios, and if it favors the softer to intermediate EOS models, it is of sufficient accuracy to inform the details of the progenitor and supernova modeling of this particular formation channel, particularly the critical mass at which the ONeMg core becomes unstable in the ECSN scenario, and the mass lost from the core during the supernovae. As more double neutron star systems are discovered in the SKA era, analyses like this one will strengthen our understanding of their possible formation pathways and corresponding consequences for astrophysics and nuclear physics.

It is of crucial importance to determine the robustness of the prediction that less than 0.015$M_{\odot}$ of mass is lost from the progenitor core, a prediction that results from a limited number of simulations. Particularly, is there a significant equation of state dependence of this quantity? Additionally, more investigation of whether the core is disrupted by runaway nuclear burning, rather than collapsing, is necessary.

We have focussed in this paper on the relation between binding energy and moment of inertia because of the prospect of a moment of inertia measurement of J0737-3039A within a decade. However, because of the universal I-Love-Q relations, a measurement of the tidal polarizability of a neutron star from the gravitational wave signal immediately prior to a binary neutron star merger to a similar degree of accuracy would also give similar constraints on the binding energy, and allow for similar discrimination of formation scenarios. Such accuracy could be obtained by Advanced LIGO for merger events within 200 Mpc. One can also use measurements of the NS radius and the relation between the binding energy and compactness to perform a similar analysis.

\section{Acknowledgments}

The authors would like to acknowledge useful discussions with Ed Brown
and James Lattimer. AWS was supported by NSF grant PHY 15-54876 and
the U.S. DOE Office of Nuclear Physics. This work used computational resources 
from the University of Tennessee and Oak Ridge
National Laboratory's Joint Institute for Computational Sciences. 
K.Y.~acknowledges support from JSPS Postdoctoral Fellowships 
for Research Abroad and NSF grant PHY-1305682.

\bibliography{bib-ecsn}
\bibliographystyle{apj}


\end{document}